\pdfoutput=1
\documentclass[11pt,a4paper]{article}
\usepackage{jcappub}

\usepackage{bm}
\usepackage{amsfonts}
\usepackage{subfigure}
\usepackage{float}

\renewcommand\[{\left[}

\newcommand{\bp}{\textbf{p}}
\long\def\dump#1{}

\begin{document}


\title{
Three flavor neutrino conversions in supernovae: Slow \& Fast instabilities
}

\author[a]{Madhurima~Chakraborty,}
\author[a,b]{Sovan~Chakraborty}

\affiliation[a]{Indian Institute of Technology, Guwahati\\
Guwahati, Assam-781039, India}
\affiliation[b]{Max-Planck-Institut f{\"u}r Physik (Werner-Heisenberg-Institut),\\
F{\"o}hringer Ring 6, 80805 M{\"unchen}, Germany}

\affiliation[a]{}
\emailAdd{sovan@iitg.ac.in}
\emailAdd{madhu176121012@iitg.ac.in}

\abstract{ Self induced neutrino flavor conversions in the dense regions of stellar core collapse are almost exclusively studied in the standard two flavor scenario. Linear stability analysis has been successfully used to understand these flavor conversions. This is the first linearized study of \textit{three flavor} fast instabilities. 
The `fast' conversions are fascinating distinctions of the dense neutrino systems. In the fast modes the collective oscillation dynamics are independent of the neutrino mass, growing at the scale of the large neutrino-neutrino interaction strength ($10^5$ km$^{-1}$) of the dense core. This is extremely fast, in comparison  to the usual `slow' collective modes driven by much smaller vacuum oscillation frequencies ($10^0$ km$^{-1}$). The three flavor analysis shows distinctive characteristics for both the slow and the fast conversions. The slow oscillation results are in qualitative agreement with the existing nonlinear three flavor studies. For the fast modes, addition of the third flavor opens up possibilities of influencing the growth rates of flavor instabilities when compared to a two flavor scenario.
}

\maketitle

\section{Introduction}
\label{sec:introduction}

Neutrino densities in extreme dense environments like collapsing supernovae (SN) cores are also extraordinarily large. In such environments the neutrinos interacting with the other background neutrinos can give rise to large refractive effects resulting in  flavor off diagonal potentials. These off diagonal potentials in turn can produce flavor conversion during the stellar core collapse\cite{Pantaleone:1992eq,Sigl:1992fn,Samuel:1993uw,Kostelecky:1993dm,Kostelecky:1994dt,Samuel:1995ri,Hannestad:2006nj,Duan:2007mv,Duan:2007fw,Fogli:2007bk,EstebanPretel:2008ni,Duan:2010bg,Duan:2010bf,Volpe:2013jgr,Vlasenko:2013fja,  Chakraborty:2016yeg}. Similar conditions may also arise in the early universe and in neutron star mergers. For long time it was understood that the driving frequency for such flavor oscillations were dependent on $\omega = \frac{\Delta m^2}{2E} \sim \mathcal{O}(1 km^{-1})$ causing flavor oscillations at few hundred kms ($\sim 10^2$ km) from the neutrinosphere.  These 
oscillations predict very rich phenomenological effects on the neutrino spectra and stellar evolution\cite{Duan:2006an,Raffelt:2007xt,Fogli:2008pt,Dasgupta:2009mg,Mirizzi:2015eza}. However, in recent times it has
been pointed out that given the suitable initial conditions of neutrino distributions deep inside the SN and given the initial triggering seeds, such self induced flavor conversions may occur even in the absence of neutrino mixing ($\Delta m^2 \sim 0$), i.e, independent of  $\omega$. Here the flavor conversion is driven by the neutrino density $\mu \sim \sqrt{2}G_F n_\nu \sim \mathcal{O}(10^5 km^{-1}) $ and may occur just above the neutrinosphere. Hence, these $\mu$ dependent oscillations are described as `fast oscillations' \cite{Sawyer:2005jk,Chakraborty:2016lct} in contrast to the $\omega$ dependent `slow' flavor conversions at radius $\sim 10^2$ km. 

Simultaneous to the fast oscillations, another very important development regarding the understanding of the `stationary' nature of the flavor evolution came into picture. It has been found that the absence of stationarity gave rise to new kind of 
instabilities, `temporal' instabilities \cite{Abbar:2015fwa,Dasgupta:2015iia}. These instabilities 
are distinctly different in comparison to the previously known flavor instabilities as the temporal instabilities are not completely `matter suppressible' \cite{Dasgupta:2015iia,Chakraborty:2016yeg}. However, these `non- matter suppressible' phases are very short and are not expected to be long enough to develop into physical flavor conversions for the slow oscillations. This picture changes very rapidly if we focus on fast oscillations. Thus the fast oscillations together with the temporal instabilities 
could generate flavor instabilities which are growing extremely fast and may avoid the problem of matter suppression. 

The techniques used to understand these phenomena are mostly based on the linear stability analysis, where exponential growths of the flavor off diagonal elements in the linearized regime signify flavor instabilities. However, combining the fast oscillations together with the nonstationary (evolution in time) and inhomogeneous (evolution in space) flavor evolution require different techniques. Since it involves both the time and spatial evolution and the related Fourier modes, it is reasonable to use the dispersion picture \cite{Izaguirre:2016gsx, Airen:2018nvp,Yi:2019hrp,Capozzi:2019lso,Martin:2019gxb}. In the dispersion picture, the imaginary Fourier modes of the disturbances
in time and space signify the instabilities, giving the most general description of the fast oscillations. 

However, all these developments in the understanding of fast oscillations, non stationarity and inhomogeneities are developed in the framework of a 2 flavor system of neutrinos. This is indeed a reasonable approximation as per many SN simulations showing an effective 2 flavor description i.e, `similar' fluxes of the mu and tau neutrinos and anti-neutrinos. This similar flux is approximated to `same' and denoted as the `$\nu_x$'. It has always remained an interesting aspect to check the validity of the `similar' flux arguments, i.e. , if the differences in the $x$ flavors are not neglected requiring a three flavor study \cite{Dasgupta:2007ws}.
Indeed, most of the three flavor studies \cite{Dasgupta:2007ws,Friedland:2010sc} pointed out that the effective 
two flavor analysis is a reasonable approximation due to the `similar' flux argument. However, we lack such a three flavor study 
in the context of the recent topics like including inhomogeneities, non-stationarity and fast oscillation.  Here we extend the techniques of the linear stability analysis and the dispersion picture of fast oscillations to a three flavor analysis. We study the impact of 
adding one extra flavour for both the slow and fast oscillations. 

For the slow oscillations the past three flavor analysis  results \cite{Friedland:2010sc, Dasgupta:2010cd} were based on numerically solving the non linear evolution. In our analysis we check if we can reproduce the same from the linearized stability analysis picture. The main result of the `slow' three flavor nonlinear analysis is that the flavor conversions due to the atmospheric and solar sector are not completely de-coupled. Thus the late onset of the solar mass squared difference which governed the flavor conversion in otherwise two flavor analysis can speed up when another extra 
flavor is added. Indeed, our linearized study  successfully reproduce these results from the past non-linear three flavor analysis. In the following, we also analyze the `fast' flavor conversion in both
time and space using the dispersion relation technique. In the 
linearized regime, the three flavor evolution decouples into three, two flavor evolutions.
These two flavor evolutions are dependent on the fluxes and densities of the different species of neutrinos. In the dispersion picture, we focus on a two beam case to realize how the addition of the extra flavor may affect the evolution. 
 We find that these three effective two flavor evolutions are sensitive to the flux differences of the $\mu$($\tau$) neutrinos and anti-neutrinos. Thus the similar to same
 approximation for the $x$ flavor fluxes may not reveal the true picture of the fast 
 conversions. These small differences can still act as a seed to influence the growth of the flavor instabilities, in turn `speeding up' or `slowing down' the fast flavor conversion in a non-linear picture. It would be interesting to see if these three flavor speeding up are really sustainable for the realistic $\nu_\mu-\nu_\tau$ flux differences in the dense SN core \cite{Bollig:2017lki}. 
 
 The paper is organized as the following. 
 In Section \ref{sec:Framework}, we discuss the framework of the linearization of neutrino-neutrino evolution
 for three flavor system of neutrinos. First, the most generalized equations are presented, then
 we outline the analysis in the context of a three flavor neutrino system. In Section \ref{sec:stability analysis}, we discuss spatial evolution of the slow modes in the linearized three flavor picture. The results are in agreement with the past non-linear three flavor analysis. In section \ref{sec:dispersion}, we present the general dispersion technique in context of the three flavor fast oscillations. 
 For an example of the system, we use the widely considered two-beam case. Finally, in Section \ref{sec:conclusion}, we
 conclude and comment on the limitation and future perspective of the present analysis.

\section{Framework}
\label{sec:Framework}
The understanding that in absence of stationarity the self induced neutrino flavor oscillations can generate very unique solutions \cite{Dasgupta:2015iia,Chakraborty:2016yeg} triggered 
the need for a picture where both the temporal and spatial degrees of freedom are treated equally. Indeed, the four vector treatment lead to the 
dispersion relation picture connecting the frequency and the wave numbers of the disturbances indicating the presence of instabilities \cite{Izaguirre:2016gsx}. In the 
following we use the framework used in the \cite{Airen:2018nvp} to describe the quantum kinetic equations for the ultra-relativistic neutrino streams. The equation of motion for flavor oscillations in absence of collisions can be written in the form 
\begin{equation}
  v^{\beta} \partial_{\beta} \rho_{\bp} = -i[H_\bp , \rho_\bp]\,.
\label{eq:flavorevol}  
\end{equation}
The neutrino velocity four-vector is defined as $v^\beta=(1,\textbf{v})$ with the three velocity vector, $\textbf{v}=\hat{\textbf{p}}=\textbf{p}/\mid\textbf{p}\mid$
for every $\textbf{p}$ mode, the usual summation convention over $\beta=0,1, 2, 3$ is implied. Thus explicitly, $v^\beta \partial_\beta$ is written as $(\partial_t+\textbf{v}\cdot \nabla )$. The Hamiltonian matrix $H_p$ includes the three components from vacuum, 
matter and the neutrino-neutrino interactions and takes the form ($E=\mid\textbf{p}\mid$)
\begin{equation}
  H_{\textbf{p}}=\frac{M^2}{2E}
  +\sqrt{2}\ G_F\,v_\beta(F_l^{\beta}+F_\nu^{\beta})\,
\end{equation}

For the vacuum oscillation term, in the initial conditions we take $M^2$ to be diagonal, i.e. in the extreme densities of neutrino neutrino interactions we ignore mixing as the matter effect
will subdue vacuum flavor conversion, 
\begin{equation}
   H^{vacuum}_p = \frac{M^2}{2E}=\frac{1}{2E}\begin{pmatrix}m_1^2&0&0\\
    0&m_2^2&0\\
    0&0&m_3^2
  \end{pmatrix}.
  \end{equation}

The matter contribution depends on the charged-current contribution of the charged leptons. In the local four-Fermion current-current description in the weak interaction basis
the matter term is $H^{matter}_p=\sqrt{2}G_F\,v_\beta F_l^\beta$, where
\begin{equation}
  F_l^\beta=\int 2\,d\textbf{p}
  \begin{pmatrix}v^\beta_e(f_{e,\textbf{p}}-\bar f_{e,\textbf{p}})&0&0\\
    0&v^\beta_\mu(f_{\mu,\textbf{p}}-\bar f_{\mu,\textbf{p}})&0\\
    0&0&v^\beta_\tau(f_{\tau,\textbf{p}}-\bar f_{\tau,\textbf{p}})
  \end{pmatrix},
\end{equation}
where $\int d\textbf{p}=\int d^3\textbf{p}/(2\pi)^3$. The charged lepton four velocity with momentum \textbf{p} is 
$v^\beta_l=(1,\textbf{v}_l)$ with $\textbf{v}_l=\textbf{p}/(\textbf{p}^2+m_l^2)^{1/2}$ and the occupation numbers are $f_{l,\textbf{p}}$ (for the antiparticle $\bar f_{l,\textbf{p}}$), $l=e,\mu,\tau$.
Similarly, the neutrino-neutrino interaction contribution to the Hamiltonian $H^{\nu\nu}_{\textbf{p}} = \sqrt{2}G_F v_\beta F_\nu^\beta$, where the neutrino flux matrix is 
$F_\nu^\beta=\int d\textbf{p}\ v^\beta (\rho_{\textbf{p}}-\bar\rho_{\textbf{p}})$  and $v^\beta = (1,\hat{\textbf{p}})$. The occupation number matrices for neutrinos and anti-neutrinos
are $\rho_{\textbf{p}}$ and $\bar\rho_{\textbf{p}}$, respectively and in general has the form
\begin{equation}
    \rho_{\textbf{p}}=
    \begin{pmatrix}
    \rho_{\textbf{p}}^{ee}&\rho_{\textbf{p}}^{e\mu}&\rho_{\textbf{p}}^{e\tau}\\ \\
    \rho_{\textbf{p}}^{\mu e}&\rho_{\textbf{p}}^{\mu\mu}&\rho_{\textbf{p}}^{\mu\tau}\\ \\
    \rho_{\textbf{p}}^{\tau e}&\rho_{\textbf{p}}^{\tau\mu}&\rho_{\textbf{p}}^{\tau\tau}\\
    \end{pmatrix}.
\end{equation}
where $\rho_p^{ll'}=\rho_p^{\nu_l\nu_{l'}}$ with l, $l'=$ e, $\mu$, $\tau$. 
Thus one may define the overall matter effect caused by both charged leptons and neutrinos, $  H^{matter}=v_{\beta}\lambda^{\beta}$, where $\lambda^\beta=(\lambda^0, \boldsymbol{\lambda}) = $ Diagonal of $\sqrt{2}\,G_F\,(F_l^\beta+F_\nu^\beta)$ $= \sqrt{2}\,G_F\,\int d\textbf{p}\,
  \Big[2\,v_l^\beta\,(f_{l,\textbf{p}}-\bar f_{l,\textbf{p}}\,)
  +v^\beta\,(\rho_{\textbf{p}}^{ll}-\bar \rho_{\textbf{p}}^{ll}\,)\Big]\,
  $ 
with $\beta = 0,..,3$. 

Note that the temporal components  $F_l^0$ and $F_\nu^0$ represent the charged lepton  and neutrino densities, respectively. The spatial parts ($\beta = 1, 2, 3$) of $F_l^\beta$ and $F_\nu^\beta$ are the corresponding current terms for charged leptons and the neutrinos. However, the usual dense neutrino systems under considerations are isotropic, hence the current terms can be neglected for most of the examples. In the absence of the current terms only the temporal components, i.e. the ordinary matter term or charged lepton density ($F_l^0$) and the neutrino density ($F_\nu^0$) are the major contributions from the matter terms (see \cite{Airen:2018nvp} for detailed discussion).

\subsection{Linearized Equations of Motion}

In compact astrophysical objects, the neutrino population are considered to be produced in flavor eigenstates making the $\rho$ matrices diagonal in the flavor basis.  Thus the flavor mixings are denoted by the growth of the off-diagonal elements. The linearized regime to begin with, the off-diagonal terms are considered to remain small and any subsequent growth will imply the presence of self-induced flavor conversion caused by neutrino-neutrino interactions. In this regard, the focus is on the off diagonal terms and they being small, it is reasonable to look
into the linearized equations of motion of the off diagonal elements ($l\neq l^\prime$) for both the neutrinos ($\rho^{l l^{\prime}}_{\textbf{p}}$) and antineutrinos ($\bar\rho^{l l^{\prime}}_{\textbf{p}}$).  

Taking into account the off-diagonals of $\rho$ upto linear order, the equations of motion are as follows 
\begin{eqnarray}
  i\,v^\beta\partial_\beta \rho_{\textbf{p}}^{e\mu}&=&
  \Bigg[\frac{m^2_1-m^2_2}{2E}
  +v_\beta(\lambda_e^\beta-\lambda_\mu^\beta)\,\Bigg]\rho_{\textbf{p}}^{e\mu}
  \nonumber\\[1ex]
  &&\kern4em{}-\sqrt{2}\,G_F\,(\rho_{\textbf{p}}^{ee}-\rho_{\textbf{p}}^{\mu\mu}\,)v^\beta
  \int d\textbf{p}'v_\beta'\,(\rho_{\textbf{p}'}^{e\mu}-\bar\rho_{\textbf{p}'}^{e\mu}\,)
  \label{eq:linoffdiagonal1}
\end{eqnarray}
\begin{eqnarray}
  i\,v^\beta\partial_\beta \rho_{\textbf{p}}^{e\tau}&=&
  \Bigg[\frac{m^2_1-m^2_3}{2E}
  +v_\beta(\lambda_e^\beta-\lambda_\tau^\beta)\,\Bigg]\rho_{\textbf{p}}^{e\tau}
  \nonumber\\[1ex]
  &&\kern4em{}-\sqrt{2}\,G_F\,(\rho_{\textbf{p}}^{ee}-\rho_{\textbf{p}}^{\tau\tau}\,)v^\beta
  \int d\textbf{p}'v_\beta'\,(\rho_{\textbf{p}'}^{e\tau}-\bar\rho_{\textbf{p}'}^{e\tau}\,)
  \label{eq:linoffdiagonal2}
\end{eqnarray}
\begin{eqnarray}
  i\,v^\beta\partial_\beta \rho_{\textbf{p}}^{\mu\tau}&=&
  \Bigg[\frac{m^2_2-m^2_3}{2E}
  +v_\beta(\lambda_\mu^\beta-\lambda_\tau^\beta)\,\Bigg]\rho_{\textbf{p}}^{\mu\tau}
  \nonumber\\[1ex]
  &&\kern4em{}-\sqrt{2}\,G_F\,(\rho_{\textbf{p}}^{\mu\mu}-\rho_{\textbf{p}}^{\tau\tau}\,)v^\beta
  \int d\textbf{p}'v_\beta'\,(\rho_{\textbf{p}'}^{\mu\tau}-\bar\rho_{\textbf{p}'}^{\mu\tau}\,)
  \label{eq:linoffdiagonal3}
\end{eqnarray}

Note that here the index $\beta$ is being summed over. In this linearized picture, the three flavor system corresponds to the three independent two flavor cases. The off diagonal elements evolve independent of
one another. Note that in principle, the evolution EOMs are also different as the vacuum and the matter terms are different for the three off-diagonal elements. 
In the subsequent discussions, we will see if these differences survive in a realistic stellar collapse and their possible effects into the overall flavor evolution.

The anti-neutrino matrices $\bar\rho_{\textbf{p}}$ have the same EOM (\ref{eq:flavorevol})  with a sign change of the vacuum oscillation term. Thus the off-diagonals 
in the anti-neutrino sector will have similar equations (\ref{eq:linoffdiagonal1}-\ref{eq:linoffdiagonal3}) with a a sign change of the vacuum oscillation term.

\subsection{Three flavor neutrino system}

According to the structure of the EOMs, the difference of the original neutrino distributions decides the flavor oscillations. The diagonal entries of the matrices drop out due to the commutation nature of the evolution equations and the off-diagonals of the occupation number matrices carry the flavor changing information. We choose the following form of the neutrino matrices involving the spectral difference,
\begin{align}
  \rho_{\textbf{p}}=&\frac{f_{\nu_e,\textbf{p}}+f_{\nu_\mu,\textbf{p}}+f_{\nu_{\tau},\textbf{p}}}{3}\,\mathbf{1}
   + \frac{f_{\nu_e,\textbf{p}}-f_{\nu_\mu,\textbf{p}}}{3}
  \begin{pmatrix}s_{\textbf{p}}&S_{1\textbf{p}}&0\\S_{1\textbf{p}}^*&-s_{\textbf{p}}&0\\0&0&0\end{pmatrix}\nonumber\\&+ \frac{f_{\nu_e,\textbf{p}}-f_{\nu_\tau,\textbf{p}}}{3}
  \begin{pmatrix}s_{\textbf{p}}&0&S_{2\textbf{p}}\\0&0&0\\S_{2\textbf{p}}^*&0&-s_{\textbf{p}}\end{pmatrix}+\frac{f_{\nu_\mu,\textbf{p}}-f_{\nu_\tau,\textbf{p}}}{3}
  \begin{pmatrix}0&0&0\\0&s_{\textbf{p}}&S_{3\textbf{p}}\\0&S_{3\textbf{p}}^*&-s_{\textbf{p}}\end{pmatrix}
\end{align}
where $s_{\textbf{p}}$ is a real number and $s_{\textbf{p}} = 1$ in the linear order. $S_{j\textbf{p}}$ are complex with $s_{\textbf{p}}^2+|S_{j\textbf{p}}|^2=1$, $j = 1,2,3$. Hence, the evolution of the off-diagonals ($j = 1,2,3$) in the linearized regime becomes
\begin{equation}
  i\,v^\beta\partial_\beta S_{j\textbf{p}}=
  \bigl(\omega_{j}+v^\beta \lambda_{j\beta}\bigr) S_{j\textbf{p}}
  -\sqrt{2}\,G_F\,v^\beta
  \int d\textbf{p}'v_\beta'\,(S_{j\textbf{p}'}g_{j\textbf{p}'}-\bar S_{j\textbf{p}'}\bar g_{j\textbf{p}'}\,).
 \label{eq:3flvoffdiagonal} 
\end{equation}
The vacuum oscillation frequencies are given by 
$\omega_{1}=\frac{(m_1^2-m_2^2)}{2E}$,
$\omega_{2}=\frac{(m_1^2-m_3^2)}{2E}$,
$\omega_{3}=\frac{(m_2^2-m_3^2)}{2E}$. The spectrum for the neutrinos are taken as $g_{1\textbf{p}}= (f_{\nu_e,\textbf{p}}-f_{\nu_\mu,\textbf{p}})$,
 $g_{2\textbf{p}}= (f_{\nu_e,\textbf{p}}-f_{\nu_\tau,\textbf{p}})$,  $g_{3\textbf{p}}= (f_{\nu_\mu,\textbf{p}}-f_{\nu_\tau,\textbf{p}})$ and for anti-neutrinos, the spectrums are given as $\bar g_{1\textbf{p}}=
(f_{\bar\nu_e,\textbf{p}}-f_{\bar\nu_\mu,\textbf{p}})$, $\bar g_{2\textbf{p}}= (f_{\bar\nu_e,\textbf{p}}-f_{\bar\nu_\tau,\textbf{p}})$, $\bar g_{3\textbf{p}}= (f_{\bar\nu_\mu,\textbf{p}}-f_{\bar\nu_\tau,\textbf{p}})$.

The effective matter terms ($\lambda_j^\beta$)  combining the effects of the charged and neutral leptons contributions are defined as, 
\begin{multline*}
\lambda_1^\beta
=(\lambda_e^\beta -\lambda_\mu^\beta)
=\sqrt{2}\,G_F \int d\textbf{p}\,
  \Big[2\,\Big(v_e^\beta\,(f_{e,\textbf{p}}-\bar f_{e,\textbf{p}}\,)-v_{\mu}^\beta\,(f_{\mu,\textbf{p}}-
  \bar f_{\mu,\textbf{p}}\,)\,\Big)\\
  + v^\beta\,\Big(\,(f_{\nu_e,\textbf{p}}-\bar f_{\nu_e,\textbf{p}}
  -f_{\nu_{\mu},\textbf{p}}+\bar f_{\nu_{\mu},\textbf{p}}\,)\,\Big)\,\Big]
\end{multline*}
\begin{multline*}
\lambda_2^\beta=(\lambda_e^\beta -\lambda_\tau^\beta)
=\sqrt{2}\,G_F\,\int d\textbf{p}\,
  \Big[2\,\Big(v_e^\beta\,(f_{e,\textbf{p}}-\bar f_{e,\textbf{p}}\,)-v_{\tau}^\beta\,(f_{\tau,\textbf{p}}-\bar f_{\tau,\textbf{p}}\,)\,\Big)\\
  +v^\beta\,\Big(\,(f_{\nu_e,\textbf{p}}-\bar f_{\nu_e,\textbf{p}}
  -f_{\nu_{\tau},\textbf{p}}+\bar f_{\nu_{\tau},\textbf{p}}\,)\,\Big)\,\Big]
\end{multline*}
\begin{multline}
\lambda_3^\beta=(\lambda_\mu^\beta -\lambda_\tau^\beta)
=\sqrt{2}\,G_F\,\int d\textbf{p}\,
  \Big[2\,\Big(v_\mu^\beta\,(f_{\mu,\textbf{p}}-\bar f_{\mu,\textbf{p}}\,)-v_{\tau}^\beta\,(f_{\tau,\textbf{p}}-\bar f_{\tau,\textbf{p}}\,)\,\Big)\\
  +v^\beta\,\Big(\,(f_{\nu_\mu,\textbf{p}}-\bar f_{\nu_\mu,\textbf{p}}
  -f_{\nu_{\tau},\textbf{p}}+\bar f_{\nu_{\tau},\textbf{p}}\,)\,\Big)\,\Big]
  \label{eq:effectivematter}
\end{multline}

To keep the description compact we follow the flavor isospin convention, i.e, antiparticles as particles with negative energy and negative occupation numbers. In this convention, the neutrino
modes are described by $-\infty<E<+\infty$ and the 
three flavor spectrums are given by,
\begin{equation*}
  g_{1E,\textbf{v}}=\begin{cases} f_{\nu_e,\textbf{p}}-f_{\nu_\mu,\textbf{p}}&\hbox{for $E>0$,}\\
  f_{\bar\nu_\mu,\textbf{p}}-f_{\bar\nu_e,\textbf{p}}&\hbox{for $E<0$,}
  \end{cases}
\end{equation*}
\begin{equation*}
  g_{2E,\textbf{v}}=\begin{cases} f_{\nu_e,\textbf{p}}-f_{\nu_\tau,\textbf{p}}&\hbox{for $E>0$,}\\
  f_{\bar\nu_\tau,\textbf{p}}-f_{\bar\nu_e,\textbf{p}}&\hbox{for $E<0$,}
  \end{cases}
\end{equation*}
\begin{equation}
  g_{3E,\textbf{v}}=\begin{cases} f_{\nu_\mu,\textbf{p}}-f_{\nu_\tau,\textbf{p}}&\hbox{for $E>0$,}\\
  f_{\bar\nu_\tau,\textbf{p}}-f_{\bar\nu_\mu,\textbf{p}}&\hbox{for $E<0$.}
  \end{cases}
\end{equation}
This choice of the spectrum makes the EOMs more compact and the integration term in equation \ref{eq:3flvoffdiagonal} simplifies to give new form,
\begin{equation}
  i\,v^\beta\partial_\beta S_{j E,\textbf{v}}=
  \bigl(\omega_{j}+v^\beta \lambda_{j\beta}\bigr) S_{jE,\textbf{v}}
  -\sqrt{2}\,G_F\,v^\beta  \int d\Gamma'\, v_\beta'\,g_{jE',\textbf{v}'}  S_{jE',\textbf{v}'}\
  \label{eq:3flvoffdiagonalspectra}
\end{equation}
where $j = 1,2,3$ and the phase-space integration is $ \int d\Gamma=\int_{-\infty}^{+\infty}\frac{E^2dE}{2\pi^2}\int \frac{d\textbf{v}}{4\pi}$.
Here, $\int d\textbf{v}$ is an integral over the unit surface, i.e., over all polar angles of $\textbf{p}$. 

The vacuum oscillation frequency changes sign for antineutrinos, i.e, for negative E. For the mass ordering we consider the usual convention 
from \cite{Chakraborty:2015tfa} and describe the energy spectrum of the neutrinos by the $\omega$ spectrum. Here, $\omega_1$ and $\omega_2$  corresponds to the solar and the 
the atmospheric sector, respectively and the $\omega_{3} = \frac{(m_2^2-m_3^2)}{2E} \simeq \omega_2$ as $\frac{\omega_{2}}{\omega_{1}}\approx 10^2$.
Therefore, in the convention of \cite{Chakraborty:2015tfa}, $\omega_{2}, \omega_{3} >0$ for inverted mass ordering (IO) and $\omega_{2}, \omega_{3} < 0$ for the normal ordering (NO).
For the solar sector, $\omega_{1} = \frac{(m_1^2-m_2^2)}{2E}<0$, thus only the solutions in the normal ordering are physical. However, in our analysis to be consistent with the other two sectors, we consider both the mass orderings even for the solar sector. 
The equations \ref{eq:3flvoffdiagonalspectra} denote the evolution of the off-diagonal terms in the matrix of densities.
These linearized evolution of the off diagonal elements are checked for any substantial growths. One may begin with a
small seed for the off diagonals and any exponential growth of these seeds denotes possibility of flavor instability. 
This is the usual stability analysis employed to understand the flavor instabilities in these dense neutrino systems.

Another more compact way of looking into the stability is the dispersion picture. The stability picture becomes
complicated if both the time-space evolution is considered. However, the equations \ref{eq:3flvoffdiagonalspectra} show 
that the complete picture requires simultaneous analysis in both space and time, hence the dispersion picture. The idea
is to look for imaginary Fourier modes of the disturbances in time ($K_0$) and space ($\textbf{K}$). The relation between 
these Fourier modes gives the dispersion relation ($D(K_{0}, \textbf{K}) = 0$).

To begin with we stick to a simpler picture of only spatial evolution and look for flavor instabilities for the slow
modes, i.e instabilities growing slowly with $\omega_i$. Here we employ the method of stability analysis for three
flavor neutrino system. Later we will look into the three flavor dispersion picture as well.

\section{Stability Analysis: Slow modes}
\label{sec:stability analysis}

In this stability picture the problem is interpreted as an instability in propagating flavor indicating the onset of the
oscillations. As already mentioned here we consider the stationary solutions, i.e., the evolution is independent of
time. Since we want to focus on the slow modes we ignore the angular distribution of the spectra ($g_{i,E,v} \equiv
g_{i,E} $). The reason to focus on such a simple scenario is that we want to compare our  three flavor results
with the existing literature \cite{Friedland:2010sc, Dasgupta:2010cd}. The previous non-linear studies in three flavor were simple in the
sense of the multi-angular treatment but had some very important  realization. It showed that if the fluxes of
the heavy lepton flavors neutrinos (usually denoted as $\nu_x$) were degenerate then the three flavor analysis can be
factorized in the two, two flavor sectors, i.e, the solar and atmospheric. The solar sector grows much slowly compare to
the atmospheric one due to the smaller $\Delta{m}^2_{ij}$ and the picture remains effectively a two flavored one. However, the 
absence of such a degeneracy will couple these two sectors and make the solar modes grow faster, thus making the three 
flavor effects important. Here we try to see if the stability analysis gives the same feature at least qualitatively. 

\subsection{Equations of Motion: Radial Evolution}

The dense neutrino systems considered in \cite{Friedland:2010sc, Dasgupta:2010cd} are stationary and only radially evolving. Moreover, the systems are taken to be isotropic and the charged lepton matter terms are neglected.

To accommodate the dense neutrino systems of  \cite{Friedland:2010sc, Dasgupta:2010cd} we consider the linearized system in accordance with \cite{Chakraborty:2014lsa} but with three flavors. To have an analogy to the \cite{Chakraborty:2014lsa} we have to understand the effective matter terms as in the present analysis the matter terms are more involved. The effective matter terms  $\lambda_{j\beta}=\lambda_{j\beta}^l+\lambda_{j\beta}^\nu$, has contributions from both the charged leptons ($l$) and neutrinos ($\nu$). The assumption of isotropy allows us to drop the current terms, i.e. $\lambda_{j1}=\lambda_{j2}=\lambda_{j3}=0$. The remaining $\lambda_{j0}$ terms denote the charged lepton and the neutrino densities. The charged lepton densities are denoted by $\lambda_{j0}^l = \lambda_j$ and the terms corresponding to the neutrino ($\lambda_{j0}^\nu$) leads to the neutrino density parameter $\mu$ and the lepton asymmetry parameter $\epsilon$. Note that in comparison to the \cite{Chakraborty:2014lsa}, there are three $\epsilon$ and $\lambda$ in the present framework. Considering the stationary and radial evolution of the system, i.e., taking $\beta=1$ only, the equations \ref{eq:3flvoffdiagonalspectra} take the form
\begin{multline}
i\partial_rS_{j,r,\omega_j, u, \phi}
=\Big[\,\omega_j+u(\lambda_j+\epsilon_j\mu)\,\Big]S_{j,r,\omega_j,u,\phi}\\
  -\mu\,\int d\Gamma_j'\,[u+u'-2\sqrt{uu'}\cos{(\phi-\phi')}]\,
  g_{j,\omega'_j,u',\phi'} S_{j,r,\omega'_j,u',\phi'}
\end{multline}

where $
  \int d\Gamma_j=\int_{-\infty}^{+\infty}d\omega_j\int_0^1 du \int_0^{2\pi} d\phi $ and $j = 1,2,3$. 
  Here, $u=\sin^2{\theta_R}$ lies in the range $0\le u \le1$ where $\theta_R$ is the emission angle relative to the radial direction of the neutrino sphere at radius R. The radial velocity at a radius r can be expressed as $v^1=v_{u,r}=(1-\frac{R^2}{r^2}u)^{1/2}$. Also note that the evolution equations have been written under the assumption of large distance approximation, i.e.,  $\frac{R}{r}<<1$, (see \cite{Chakraborty:2014lsa} for more detail). 
  
  In this analysis the main difference is the three flavors of charged leptons and neutrinos, contrary to the usual two flavor analysis. Thus in principle there are three matter densities and three lepton asymmetry parameters,
\begin{equation*}
    \lambda_1=\frac{\sqrt{2}G_FR^2}{2r^2}\big[2((n_e-\bar{n}_e)-(n_\mu-\bar{n}_\mu))\big], ~~~ \lambda_2=\frac{\sqrt{2}G_FR^2}{2r^2}\big[2((n_e-\bar{n}_e)-(n_\tau-\bar{n}_\tau))\big],
    \end{equation*}
\begin{equation}
\lambda_3=\frac{\sqrt{2}G_FR^2}{2r^2}\big[2((n_\mu-\bar{n}_\mu)-(n_\tau-\bar{n}_\tau))\big],
\label{eq:lambda}
\end{equation}
where $n_e, n_\mu, n_\tau$ are the net electron, muon and tauon densities respectively. The effective neutrino density 
is defined as ,
\begin{equation}
    \mu= \sqrt{2}G_F\frac{\Big[\,2(N_{\nu_e}+N_{\bar{\nu}_e})-(N_{\nu_\tau}+N_{\bar{\nu}_\tau}+N_{\nu_\mu}+N_{\bar{\nu}_\mu})\,\Big]}{4\pi r^2}\frac{R^2}{2r^2},
\end{equation}
 here $N_{\nu_e},N_{\bar{\nu}_e}$ are the total number of electron neutrinos and antineutrinos and similarly for other flavors. Therefore lepton asymmetry parameters are 
\begin{eqnarray*}
    \epsilon_1=\frac{\int d\Gamma_1\,(f_{\nu_e}-f_{\nu_\mu})}{N}=\frac{(N_{\nu_e}-N_{\bar{\nu}_e})-(N_{\nu_\mu}-N_{\bar{\nu}_\mu})}{N},\\
    \epsilon_2=\frac{\int d\Gamma_2\,(f_{\nu_e}-f_{\nu_\tau})}{N}=\frac{(N_{\nu_e}-N_{\bar{\nu}_e})-(N_{\nu_\tau}-N_{\bar{\nu}_\tau})}{N},
\end{eqnarray*}    
\begin{equation}
    \epsilon_3= \frac{\int d\Gamma_3\,(f_{\nu_\mu}-f_{\nu_\tau})}{N}=\frac{(N_{\nu_\mu}-N_{\bar{\nu}_\mu})-(N_{\nu_\tau}-N_{\bar{\nu}_\tau})}{N},
\label{eq:epsilon}
\end{equation}

where $N=\Big[\,2(N_{\nu_e}+N_{\bar{\nu}_e})-(N_{\nu_\tau}+N_{\bar{\nu}_\tau}+N_{\nu_\mu}+N_{\bar{\nu}_\mu})\,\Big]$.\\
 In order to find the unstable modes, we assume the solutions of the linearized equations to be of the form $S_{j,r,\omega_j,u,\phi}=Q_{j,\omega_j,u,\phi}e^{-i\Omega_jr}$,
leading to the eigenvalue equations in $Q_j$s. 
\begin{equation}
    \Big[\omega_j+u(\lambda_j+\epsilon_j\mu)-\Omega_j\,\Big]Q_{j,\omega_j,u,\phi} = \mu\,\int d\Gamma_j'\,[u+u'-2\sqrt{uu'}\cos{(\phi-\phi')}]\,g_{j,\omega_j} Q_{j,\omega_j,u,\phi}\,
\label{eq:lineigenvalue}    
\end{equation}
The eigenvalues in principle are complex ($\Omega_j=\gamma_j+i\kappa_j, j=1,2,3$). Any non-zero imaginary part 
($\kappa_j \neq 0$) would signify unstable modes with the growth rates $\kappa_j$. Notice that as we are focusing on 
slow collective modes the spectrum ($g_{j,\omega}$) is only energy dependent. There is no angular dependence ($u,\phi$) 
on spectrum as necessary for finding fast flavor oscillation modes. \\
Here, we consider monochromatic neutrinos, i.e., delta functions with some fixed energy $\omega_j=\pm\omega_{j'}$, 
\begin{equation}
g_{j,\omega_j}=\Big(1+\frac{\epsilon_j}{2}\Big)\,\delta(\omega_j-\omega'_j)-\Big(1-\frac{\epsilon_j}{2}\Big)\,\delta(\omega_j+\omega'_j)\,,
\label{eq:spectramono}
\end{equation}
where $\epsilon_j$ is the spectral asymmetry.\\
Now the focus on a `realistic' scenario of the usual SN environment. The usual practice is to assume that all three heavy lepton flavour neutrinos (i.e., $\nu_{\mu},\bar\nu_{\mu}, \nu_{\tau},\bar\nu_{\tau}$) have identical number density and spectra. However, our analysis shows that the minimal requirement is that the number of mu and tau type neutrinos to be identical to their corresponding antineutrinos, i.e., $N_{\nu_\mu}=N_{\bar{\nu}_\mu}$ and $N_{\nu_\tau}=N_{\bar{\nu}_\tau}$. Moreover, the usual supernova scenario, muons and tauons are considered to be absent or negligible\footnote[1]{Recent SN simulations \cite{Bollig:2017lki} show the possibility of presence of muons in the late accretion phase. See sections \ref{subsec:numericaleg.} and \ref{sec:conclusion} for more details.}.
Thus, from equations \ref{eq:lambda} and \ref{eq:epsilon}, 
\begin{equation*}
\lambda_1 = \lambda_2 = \lambda ~;~ \lambda_3 = 0 ~~ and ~~ 
\epsilon_1 = \epsilon_2 = \epsilon ~;~  \epsilon_3 = 0\,.
\end{equation*} 
Now, the evolution equations \ref{eq:lineigenvalue} become,
\begin{equation}
    \Big[\omega_1+u \bar\lambda-\Omega_1\,\Big]Q_{1,\omega_1,u,\phi} = \mu\,\int d\Gamma_1'\,[u+u'-2\sqrt{uu'}\cos{(\phi-\phi')}]\,g_{1,\omega_1}Q_{1,\omega_1,u,\phi}\,,
    \label{eq:SNeigenevol1}
\end{equation}
\begin{equation}
    \Big[\omega_2+u \bar\lambda -\Omega_2\,\Big]Q_{2,\omega_2,u,\phi} = \mu\,\int d\Gamma_2'\,[u+u'-2\sqrt{uu'}\cos{(\phi-\phi')}]\,g_{2,\omega_2}Q_{2,\omega_2,u,\phi}\,,
    \label{eq:SNeigenevol2}
\end{equation}
\begin{equation}
    \Big[\omega_2-\Omega_3\,\Big]Q_{3,\omega_3,u,\phi} = \mu\,\int d\Gamma_2'\,[u+u'-2\sqrt{uu'}\cos{(\phi-\phi')}]\,g_{3,\omega_3}Q_{3,\omega_3,u,\phi}\,,
    \label{eq:SNeigenevol3}
\end{equation}

where, $\bar\lambda = \lambda+\epsilon\mu$.
In order to solve the equations \ref{eq:SNeigenevol1} - \ref{eq:SNeigenevol3}, we adopt a similar mechanism as done in 
\cite{Raffelt:2013rqa, Chakraborty:2014lsa, Chakraborty:2015tfa} for the two flavor case. We can reasonably employ the 2 flavor method as the three flavor
evolution is factorized into three independent two flavor evolution. In the linearized picture this independent
evolution of the three off diagonal elements makes the job simple and give qualitative understanding of the onset of 
the evolution. In the nonlinear evolution all the three modes will not be independent and onset in one sector will speed
up the onset in other sectors due their coupled nature of evolution.\\
In particular, to make things simpler we consider the scenario where $(\bar{\lambda}=0)$. 
The main impact of this matter term is to `suppress' the growth of the instability. 
However, our focus is to 
understand the dynamics of the three off diagonal elements in this linear regime
and to check if they qualitatively give the same picture obtained in the simple nonlinear 
analysis \cite{Dasgupta:2010cd}.
Thus the instability conditions in each of the 2 flavor systems are \cite{Raffelt:2013rqa, Chakraborty:2014lsa}, 
\begin{equation}
(I_{j1}-1)^2=I_{j0} I_{j2}
\quad\hbox{and}\quad
I_{j1}=-1\,,
\label{eq:instabcond}
\end{equation}
where 
\begin{equation}
    I_{1n}=\mu\int d\omega_1\,du\,\frac{u^ng_{1,\omega_1}}{\omega_1-\Omega_1}\,,
    \label{eq:instabinteg1}
\end{equation}
\begin{equation}
    I_{2n}=\mu\int d\omega_2\,du\,\frac{u^ng_{2,\omega_2}}{\omega_2-\Omega_2}\,,
    \label{eq:instabinteg2}
\end{equation}
and 
\begin{equation}
    I_{3n}=\mu\int d\omega_2\,du\,\frac{u^ng_{3,\omega_2}}{\omega_2-\Omega_3}\,.
    \label{eq:instabinteg3}
\end{equation}
Note that the equations \ref{eq:instabinteg1} - \ref{eq:instabinteg3} has a dependency on the asymmetry parameter $\epsilon$ through the spectrum $g_{j,\omega_j}$

\subsection{Results: Three flavor instability plots}

Solving the equations \ref{eq:instabcond} and using the spectrum of \ref{eq:spectramono} leads to a quartic 
equation for the first block and a quadratic equation for the second one in the instability conditions. 
The general form of the quartic and quadratic equations are as follows,
\begin{equation}
\frac{-\mu^2}{3}\Bigg(\frac{1+\frac{\epsilon_j}{2}}{\omega_j'-\Omega_j}+
\frac{1-\frac{\epsilon_j}{2}}{\omega_j'+\Omega_j}\Bigg)^2
+\Bigg[-1+\frac{\mu}{2}\Bigg(\frac{1+\frac{\epsilon_j}{2}}{\omega_j'-\Omega_j}+
\frac{1-\frac{\epsilon_j}{2}}{\omega_j'+\Omega_j}\Bigg)\Bigg]^2=0
\label{eq:quatric}
\end{equation}
\begin{equation}
1+\frac{\mu}{2}\Bigg(\frac{1+\frac{\epsilon_j}{2}}{\omega_j'-\Omega_j}
+\frac{1-\frac{\epsilon_j}{2}}{\omega_j'+\Omega_j}\Bigg)=0
\label{eq:quadratic}
\end{equation}
where $j=1,2,3$. 
\begin{figure}
    \centering
    \includegraphics[width=0.556\textwidth]{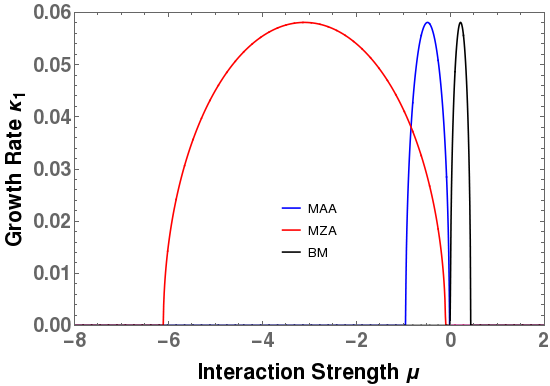}
    \vskip6pt
    \hskip8pt
    \includegraphics[width=0.55\textwidth]{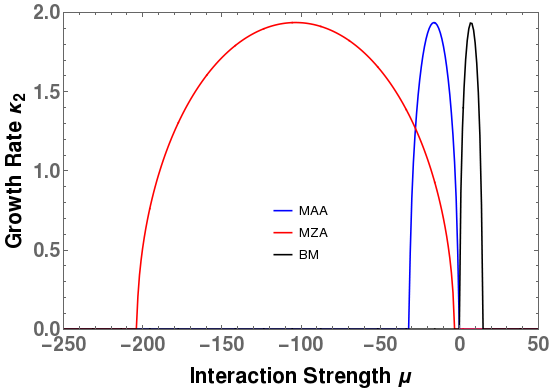}
    \vskip10pt
    \hskip18pt
    \includegraphics[width=0.557\textwidth]{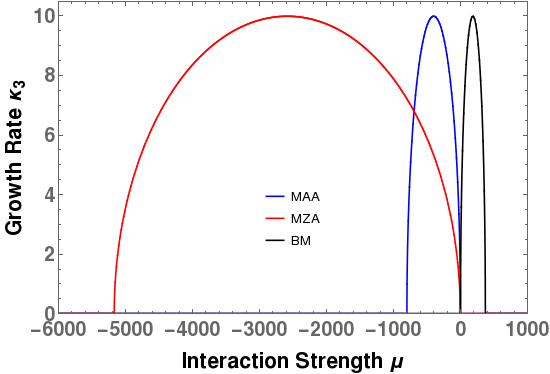}
    \caption{\footnotesize Growth rate $\kappa$ versus interaction strength $\mu$ for the unstable modes in the homogeneous (k=0) case for the evolution of $S_1$(First Panel), $S_2$(Second Panel), $S_3$(Third Panel). Here, we take $\epsilon_1=\epsilon_2=\epsilon=0.5, \epsilon_3=10^{-1}$, $\omega'_1=0.015,\omega'_2=0.5, \omega'_3=0.5$. The first block in equation \ref{eq:instabcond} yields the bimodal and MZA instability (black and red curves) and the second block yields multi-azimuth-angle instability (blue curve).}
    \label{fig:kappa-mu}
\end{figure}

Given the values of $\omega'_j$ and $\epsilon$, one may look for the imaginary solutions of these equations. The presence of imaginary solutions for the equations \ref{eq:quatric} indicates the presence of two kinds of instabilities, namely Bimodal (BM) and Multi Zenith Angle (MZA) instability whereas, for equations \ref{eq:quadratic}, it shows the Multi Azimuthal Angle (MAA) instability \cite{Chakraborty:2015tfa}. 

To probe the solutions numerically, we estimate the values of $\omega_j'$ and $\epsilon$ similar to the supernova
case. Here, $\omega_1'$, $\omega_2'$ and $\omega_3'$ depends on the relevant mass square values. We take $\frac{\omega_1'}{\omega_2'}=\frac{7.3\times10^{-5}\,eV^2}{2.4\times10^{-3}\,eV^2}\simeq3\times10^{-2}$ \cite{Capozzi:2018ubv}. We use
$\omega_2'= 0.5 $ and then $\omega_1'$ becomes $0.015$. For the asymmetry parameter we take $\epsilon=0.5$. Thus the evolution of the first two off diagonal elements (S1 and S2) are similar but driven by two 
different scales due to the difference in solar and atmospheric mass square difference.\\ However, for the 
third off diagonal element the evolution is trivial for the spectrum $g_{3,\omega_3} = 0$ or $\epsilon_3=0$. In case the 
spectrum of the $\mu$ and $\tau$ neutrinos are same then $g_{3,\omega_3} = 0$ thus the evolution effectively
factorizes in $2$ two flavor evolutions. The solar and the atmospheric sector would decouple. This scenario is shown in the upper two panels
of figure \ref{fig:kappa-mu}. 

From the solutions of equations \ref{eq:quatric} and \ref{eq:quadratic}, it is possible to understand the variation of the maximum growth rate $\kappa^{max}$ and the corresponding interaction strength $\mu^{onset}$ with the chosen vacuum frequency $\omega_j'$ and the asymmetry parameter $\epsilon$. The relation between them comes out to be $\kappa^{max}=\sqrt{(2/\epsilon)^2-1}$ \cite{Chakraborty:2014lsa}, where $\kappa^{max}$ is expressed in the units of vacuum frequency $\omega_j'$. 
For our chosen numerical values of $\omega_j'$ and $\epsilon$ we find $\mu^{onset}_2>\mu^{onset}_1$ and correspondingly $\kappa_2^{max}>\kappa_1^{max}$. In particular, $\kappa_1^{max}=0.06$ and $\kappa_2^{max}=1.9$ is in excellent agreement with
the upper two panels of figure \ref{fig:kappa-mu},.

For all the cases BM, MAA and MZA instabilities are present. The BM happens for positive $\mu$, i.e, IO and the 
MAA and MZA happens for the negative $\mu$, i.e, NO. From the figures it is clear that for S1 (solar sector)
the instability happens at much smaller $\mu$ in comparison to the S2 (atmospheric sector). Smaller $\mu$ 
would mean larger radius. Thus the instabilities for the atmospheric sector happen at a smaller radius, 
closer to the core. Due to the smaller mass squared difference the instabilities slow down in the solar sector. In fact, the neutrino density $\mu$ with its $1/r^4$ dependence would be too small to sustain the 
solar instability to any substantial flavor conversion.

However, the realistic SN case might be different than this idealistic situation, even a small $\nu_{\mu}-\nu_{\tau}$
asymmetry would trigger the third off-diagonal element S3. In that case, equations \ref{eq:quatric} and \ref{eq:quadratic} would also hold for $S_3$ and result in a similar evolution in the $\nu_\mu-\nu_\tau$ sector as well. This has been shown in the lower most panel in figure \ref{fig:kappa-mu}. We have used $\omega_3'=\omega_2'=0.5$ and $\epsilon_3 = 10^{-1}$. Clearly, even for this small asymmetry S3
grows much faster than the other two off-diagonals. Here, the maximum growth rate comes out to be $\kappa_3^{max}=9.987$. Thus the linear evolution is decoupled in these three two flavor evolutions. Hence, the non linear evolution will not decouple the solar and atmospheric sector as the $\nu_{\mu}-\nu_{\tau}$ sector will act as the coupling factor and speed up the conversion \cite{Dasgupta:2010cd}. 

Thus the simple stability analysis in the linear regime for the three flavor system helps in
gaining analytic understanding of the system. We reproduce the existing understanding of the three flavor system of collective neutrino evolution in SN environment. The system we 
considered is a simple one, without temporal evolution, homogeneous and no angular 
distributions for the spectrum, i.e, no fast oscillations. In the following, we discuss the 
three flavor fast conversion in the dispersion picture for evolution in both time and space.

\section{Dispersion Relation: Fast Oscillation}
\label{sec:dispersion}
One common assumption in the discussions of dense neutrino systems was stationary 
solutions for the flavor evolution. However, the breaking of stationarity came out 
with surprising development of new instabilities, termed as `temporal instabilities'. 
Around the same time it was also found that there is a complete different class
of instability connected to the angular distributions of different flavors. In
particular, the angular  ($u, \phi$) dependence of the spectrum ($g_{j, \omega_{j}, u,
\phi}$) is the origin to these new instabilities. These new class of instabilities 
grow very fast, growth rates comparable to the neutrino density ($\mu$), termed 
as fast instabilities resulting in fast conversions! These instabilities are markedly 
different from the usual `slow conversions' growing at the rate of 
the vacuum frequencies ($\omega_j$). In this regard, it is necessary 
that both the temporal and spatial instabilities are studied together to
understand the true nature of the fast oscillations. Hence, the approach of  
dispersion relation, i.e, the relation between frequencies ($K_j^0,$) and the
wave numbers ($\textbf{K}_j$) of the temporal and spatial disturbances,
respectively, indicates the presence of instability \cite{Izaguirre:2016gsx}.

In particular, any imaginary $K_j^0$ for real $\textbf{K}_j$ (i.e. ) and vice versa will denote exponential
growth of the temporal and spatial disturbances. Here, it is worth noting that
the dispersion relation approach can also be used for slow conversion. 
Here, in order to understand the fast flavor conversions for a three flavor system 
we keep it simple and study the dispersion relation only for the fast instabilities 
without any coupling to slow conversion.\\
Consider the space-time dependent solutions for the three ($j=1,2,3$) off diagonal elements
\begin{equation}
    S_{j,\,\Gamma,\,r}=Q_{j,\,\Gamma,\,K}\,e^{-i(K_j^0 t-\textbf{K}_{j}\cdot\textbf{r})}\,,\,
\end{equation}
where $\Gamma = (E, \textbf{v})$, $r=(t,\textbf{r})$ and $K_j=(K_j^0,\textbf{K}_j)$. Inserting the above ansatz in equations \ref{eq:3flvoffdiagonalspectra}, we get
\begin{equation}
  \Big[ v_\beta(K^\beta_j-\lambda_j^\beta)-\omega_j\Big] Q_{j,\,\Gamma,\,K}=v_\beta A^\beta_{j,\,K},
  \qquad\hbox{with}\qquad
  A^\beta_{j,\,K}=-\int d\Gamma^{'}\, v^{'\beta}\,g_{j,\,\Gamma^{'}}\, Q_{j,\,\Gamma^{'},\,K}.
\label{eq:3flveqdispersion}
\end{equation}
Provided the quantity $ v_\beta(K^\beta_j-\lambda^\beta_j)-\omega_j\neq0$ ($j=1,2,3$), possible solutions for equations \ref{eq:3flveqdispersion} will take the form
$  Q_{j,\,\Gamma,\,K}=\frac{v_\beta A^\beta_{j,\,k}}{v_\beta(K^\beta_j-\lambda_j^\beta)-\omega_j}$. Thus the evolution equations take the form 
\begin{equation}
  v_\beta A^\beta_{j,\,k}=-v^\beta A^\alpha_{j,\,K}\int d\Gamma'\,g_{j,\,\Gamma'}\,
  \frac{v'_\beta v'_\alpha}{v'_\gamma (K^\gamma_j-\lambda^\gamma_j)-\omega_j}\,.
\end{equation}

In the following the equations are written in a compact form with the help of the `polarization' tensor $\Pi_{j, K}^{\alpha\beta}$.
There are total three sets of equations one for each $j$,
\begin{equation*}
v_\beta\,\Pi_{1,K}^{\alpha\beta}A_{1,K, \alpha}=0
\qquad\hbox{with}\qquad
\Pi_{1,K}^{\alpha\beta}=\eta^{\alpha\beta}+\int d\Gamma'\,g_{1,\,\Gamma'}\,
\frac{v^\alpha v^\beta}{v'_\gamma(K^\gamma_1-\lambda^\gamma_1) -\omega_1}\,,
\end{equation*}
\begin{equation}
v_\beta\,\Pi_{2,K}^{\alpha\beta}A_{2,K, \alpha}=0
\qquad\hbox{with}\qquad
\Pi_{2,K}^{\alpha\beta}=\eta^{\alpha\beta}+\int d\Gamma'\,g_{2,\,\Gamma'}\,
\frac{v^\alpha v^\beta}{v'_\gamma(K^\gamma_2-\lambda^\gamma_2) -\omega_2}\,,
\label{eq:polarizationdefn}
\end{equation}
\begin{equation*}
~~~~~~v_\beta\,\Pi_{3,K}^{\alpha\beta}A_{3,K, \alpha}=0
\qquad\hbox{with}\qquad
\Pi_{3,K}^{\alpha\beta}=\eta^{\alpha\beta}+\int d\Gamma'\,g_{3,\,\Gamma'}\,
\frac{v^\alpha v^\beta}{v'_\gamma(K^\gamma_3-\lambda^\gamma_3) -\omega_3}\,,
\end{equation*}
where $\eta^{\alpha\beta}=$\,\,diag(+,-,-,-) is the metric tensor. The equations hold for any $v_\beta$ leading to four independent equations $\Pi_{j,K}^{\alpha\beta}A_{j,K, \alpha}=0$, where j=1,2,3  corresponding to each evolution equation. Non-trivial solutions of these equations give the resultant dispersion relation connecting the frequency ($K^0_j$) and the wave numbers ($\textbf{K}_j$),
\begin{equation}
  {\rm det}\,\Pi^{\alpha\beta}_{j,K}=0\,,
\end{equation}
 In our simplified system, these dispersion relations depend on the neutrino flavor spectrum $g_{j, \Gamma}$ and the vacuum oscillation frequency $\omega_j$. In the following, we focus our analysis to the problem of fast flavor oscillations dominated by the neutrino densities rather than the vacuum frequency $\omega_j$.

\subsection{Fast Flavor Limit}

To study the fast modes, we consider $M^2=0$, i.e., $\omega_j \to 0$. In this case, the EOM no longer depends on E and we deal with the angular modes only. Therefore, the energy integrals in \ref{eq:polarizationdefn} can be performed on the distribution functions alone. So, we define
\begin{eqnarray}
  G_{1,\textbf{v}}=\int_{-\infty}^{+\infty}\frac{E^2 dE}{2\pi^2}\,g_{1,E,\textbf{v}}
  =\sqrt{2}\,G_F\int_{0}^{\infty}\frac{E^2 dE}{2\pi^2}\,
  \bigl(f_{\nu_e,\textbf{p}}-f_{\bar\nu_e,\textbf{p}}-f_{\nu_\mu,\textbf{p}}+f_{\bar\nu_\mu,\textbf{p}}\bigr) \, ,  \nonumber\\
  G_{2,\textbf{v}}=\int_{-\infty}^{+\infty}\frac{E^2 dE}{2\pi^2}\,g_{2,E,\textbf{v}}
  =\sqrt{2}\,G_F\int_{0}^{\infty}\frac{E^2 dE}{2\pi^2}\,
  \bigl(f_{\nu_e,\textbf{p}}-f_{\bar\nu_e,\textbf{p}}-f_{\nu_\tau,\textbf{p}}+f_{\bar\nu_\tau,\textbf{p}}\bigr) \, ,  \\
   G_{3,\textbf{v}}=\int_{-\infty}^{+\infty}\frac{E^2 dE}{2\pi^2}\,g_{3,E,\textbf{v}}
  =\sqrt{2}\,G_F\int_{0}^{\infty}\frac{E^2 dE}{2\pi^2}\,
  \bigl(f_{\nu_\mu,\textbf{p}}-f_{\bar\nu_\mu,\textbf{p}}-f_{\nu_\tau,\textbf{p}}+f_{\bar\nu_\tau,\textbf{p}}\bigr) \, ,  \nonumber
\label{eq:fastflvangdist}  
\end{eqnarray}

with $\textbf{p}=E\textbf{v}$. For each $j$, the polarization tensors $\Pi_{j,K}^{\alpha\beta}$  become, 
\begin{equation}
  \Pi_{j,K}^{\alpha\beta}=\eta^{\alpha\beta}+\int \frac{d\textbf{v}}{4\pi}\,G_{j,\,\textbf{v}}\,
  \frac{v^\alpha v^\beta}{v_\gamma(K^\gamma_j-\lambda^\gamma_j)}\,
\end{equation}
The denominator $v_\gamma(K^\gamma_j-\lambda^\gamma_j)$ is takes the form $(K^0_j-\lambda_j^0)-\textbf{v}\cdot(\textbf{K}_j-\boldsymbol{\lambda_j})$
and solved to form the dispersion relation $D (K^0_j, \textbf{K}_j) = 0$. 
\subsection{Two-Beam Case}
In the following, we use the widely used `simplest yet non-trivial' example \cite{Izaguirre:2016gsx, Capozzi:2017gqd} of the two beam case. 
With two angle modes representing two zenith ranges, it is termed as the `two beam' neutrino model. In particular, we use the general 
formalism of the of the classification of instabilities \cite{Capozzi:2017gqd},
\begin{equation}
    G_{j,\textbf{v}}=g'_{j}\frac{4\pi\,\delta(\textbf{v}-\textbf{v}_1)}{1-\textbf{v}_1\cdot\textbf{v}_2} + g''_{j}\frac{4\pi\,\delta(\textbf{v}-\textbf{v}_2)}{1-\textbf{v}_1\cdot\textbf{v}_2}
\label{eq:2beamangdist}    
\end{equation}
The equations \ref{eq:3flvoffdiagonal} govern the evaluation of the off diagonal elements and in the fast flavor limit they are written as,
\begin{equation}
  i\,v^\beta\partial_\beta S_{1\textbf{v}}=
  \bigl(v^\beta \lambda_{1\beta}\bigr) S_{1\textbf{v}}
  -v^\beta  \int \frac{d\textbf{v}}{4\pi}\, v_\beta'\,G_{1,\textbf{v}'}  S_{1\textbf{v}'}\,,
\end{equation}
\begin{equation}
  i\,v^\beta\partial_\beta S_{2\textbf{v}}=
  \bigl(v^\beta \lambda_{2\beta}\bigr) S_{2\textbf{v}}
  -v^\beta  \int \frac{d\textbf{v}}{4\pi}\, v_\beta'\,G_{2,\textbf{v}'}  S_{2\textbf{v}'}\,,
\end{equation}
\begin{equation}
  i\,v^\beta\partial_\beta S_{3\textbf{v}}=
  \bigl(v^\beta \lambda_{3\beta}\bigr) S_{3\textbf{v}}
  -v^\beta  \int \frac{d\textbf{v}}{4\pi}\, v_\beta'\,G_{3,\textbf{v}'}  S_{3\textbf{v}'}\,,
\end{equation}

where the angular distributions $G_{j,\textbf{v}}$ is given above. For our system we assume azimuthal symmetry with respect to 
the propagation direction (say z-direction) and translational symmetry in the transverse directions, hence the relevant dynamics involves
only one space coordinate $z$ or its conjugate $K_z$ (henceforth  denoted as $K$). 
In the formalism of \cite{Capozzi:2017gqd}, find the general dispersion relation for $K^0_j$ and $K_j$ ($j=1,2,3$),
\begin{equation}
    (K^0_j - \lambda_j^0 - v_1 K_j + \lambda_j)(K^0_j - \lambda_j^0 - v_2 K_j + \lambda_j) = g_j'\,g''_j.
    \label{eq: disprelation}
\end{equation}
Here, $v_1, v_2$ and $\lambda_j$ are the projections of the neutrino velocities and matter density along the z direction, respectively. The dispersion relation from the equation \ref{eq: disprelation} can be solved either in terms of $K^0_j$ or $K_j$ and the solutions are:
\begin{equation}
K^0_j = \frac{1}{2}\Bigg[2\lambda_j^0 + (v_1 +v_2)(K_j - \lambda_j) \pm \sqrt{4g'_jg_j'' + (K_j - \lambda_j)^2(v_1 - v_2)^2}\,\Bigg]
\end{equation}
\begin{equation}
K_j = \frac{1}{2v_1v_2}\Bigg[2 \lambda_j v_1 v_2+ (v_1 + v_2)(K^0_j - \lambda_j^0)\pm\sqrt{ 4 g'_j g_j'' v_1 v_2 + (K^0_j -\lambda_j^0)^2(v_1 - v_2)^2}\,\Bigg]
\end{equation}
Given the angular distribution ($v_1, v_2, g_j'\,g''_j$) and the density ($\lambda$) one may find the dispersion relation $D (K^0_j, \textbf{K}_j) = 0$.

\subsection{Numerical example}
\label{subsec:numericaleg.}
To exhibit the 3 flavor dispersion relations we consider the following numerical example.
Usually the SN like examples are simplified with the argument that the fluxes of $\nu_\mu$ and $\nu_\tau$ are almost equal to their corresponding antineutrino fluxes. In our case, from the definition of the distribution
functions in equations \ref{eq:fastflvangdist}, this would result in $G_{3,\textbf{v}}=0$ and $G_{1,\textbf{v}}=G_{2,\textbf{v}}$. Thus effectively there will be only one 
dispersion relation, similar to that of two flavor evolution. Hence, the limit $F_{\nu_\mu} = F_{\bar\nu_\mu}$ and $F_{\nu_\tau} = F_{\bar\nu_\tau}$ would effectively 
reproduce the two flavor results. 
However, for our purpose we consider the simplest two beam model where each beam consists of all the three flavors of neutrino.
Therefore, to separate the three flavor effects we consider a slight difference in the fluxes, i.e, the $g'_j$ and $g''_j$ fluxes are not same for different
$j$'s. Another interesting fact is that, the values of the matter terms in equations \ref{eq:lambda} are constant but can be different \cite{Bollig:2017lki} for the three evolution
equations as they involve the densities of the charged leptons and neutrinos (see equation \ref{eq:effectivematter}). Thus the matter term can not be rotated away by redefining the $K^0_j$ and $K_j$, unlike the previous studies
\cite{Capozzi:2017gqd}. Here, that would mean different rotations for the three different evolution equations as a result of which the new `homogeneous' modes \cite{Airen:2018nvp} will not be same for the three different off diagonals. Hence we do not drop the matter terms.

For our example, we consider the effects of the heavy charged leptons ($n_\mu, n_\tau$) to be secondary compared to the order of
the asymmetry in matter densities of electrons-positrons. This is in analogy to a SN like situation. Even if there are some muons present in the system, the presence of $\tau$ are considered to be extremely negligible. Thus the
asymmetry between the different charged leptons and their antiparticles will have the hierarchy, $n_e-n_{\bar{e}} > n_\mu - \bar{n}_\mu \gg n_\tau - \bar{n}_\tau$ and similarly for neutrinos. In particular, we consider the following example, $\lambda_1^\gamma=0.9\lambda_2^\gamma$ and $\lambda_3^\gamma=0.1\lambda_2^\gamma$, where $\lambda_j^\gamma=(\lambda_j^0,\boldsymbol{\lambda}_j)$. We choose the numerical values analogous to the early accretion phase \cite{Chakraborty:2011gd,Chakraborty:2011nf} where the neutrino densities are similar to the lepton densities. Here, we also take into consideration the neutrino and lepton current terms (spatial parts of $\lambda_j^\gamma$) unlike the previous case of stability analysis where they have been neglected. In particular, we consider $\lambda_2^0= \lambda_2=10$ thus $\lambda_1^0= \lambda_1=9$ and $\lambda_3^0= \lambda_3=1$.

\begin{figure}[H]
\centering
\hbox{\includegraphics[width=0.5\textwidth]{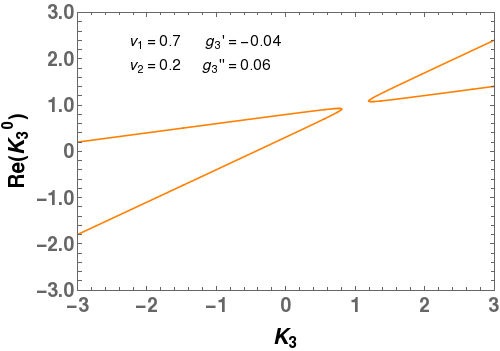}\hspace{0.5cm}\includegraphics[width=0.5\textwidth]{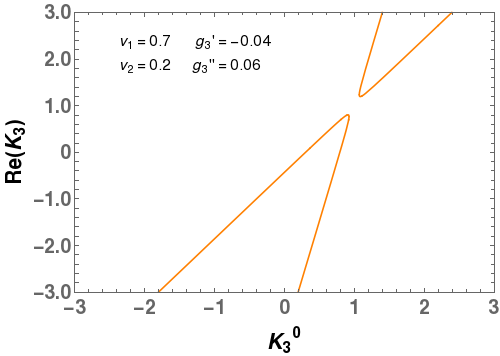}}
\hbox{\includegraphics[width=0.5\textwidth]{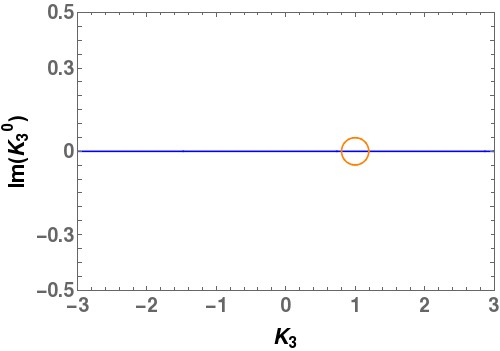}\hspace{0.5cm}\includegraphics[width=0.5\textwidth]{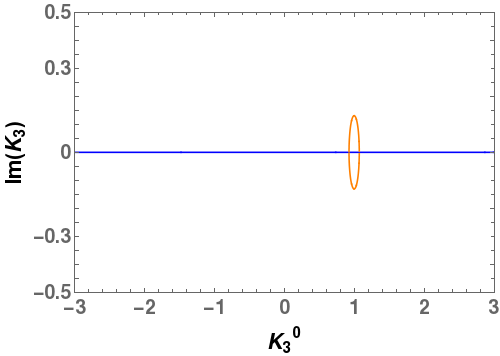}}
\hbox{\includegraphics[width=0.5\textwidth]{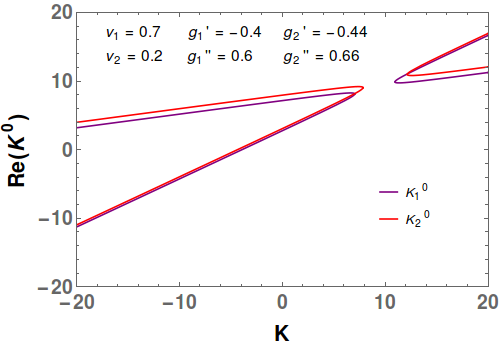}\hspace{0.5cm}\includegraphics[width=0.5\textwidth]{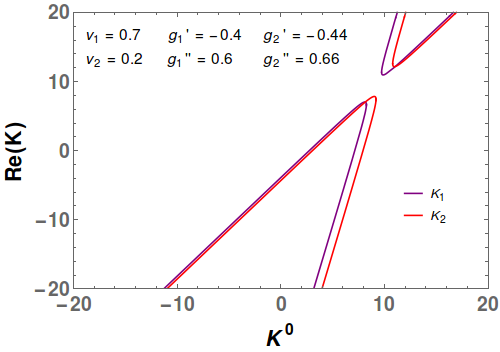}}
\hbox{\includegraphics[width=0.5\textwidth]{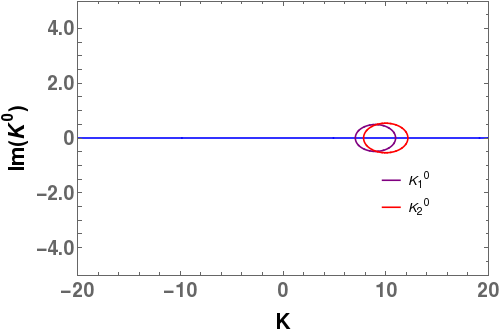}\hspace{0.5cm}\includegraphics[width=0.5\textwidth]{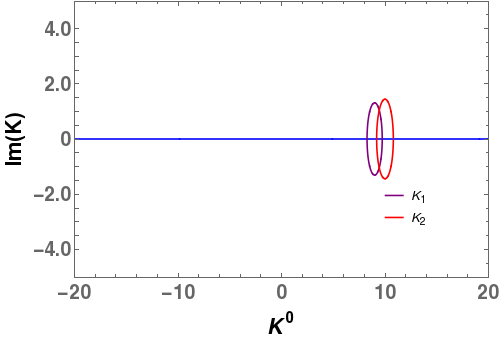}}
\caption{Example of solutions of the dispersion relation (4.12) for a convectively unstable case. The upper two panels are for j=3 and lower two panels for j=1,2.  Here $v_1=0.7,v_2=0.2$. 
$g'_3=-0.04,g''_3=0.06$ and $g'_1=-0.4,g''_1=0.6$, $g'_2=-0.44,g''_2=0.66$.} 
\label{fig:dispersion2}
\end{figure}

\begin{figure}[H]
\centering
\hbox{\includegraphics[width=0.5\textwidth]{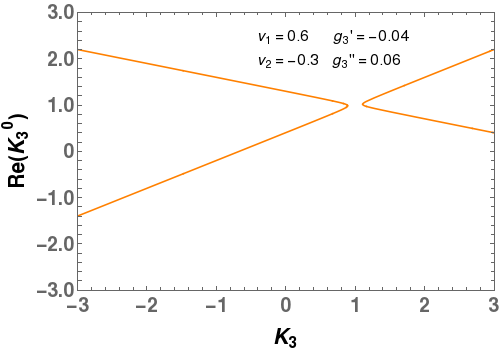}\hspace{0.5cm}\includegraphics[width=0.5\textwidth]{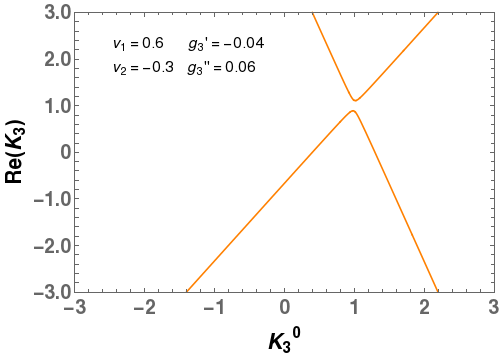}}
\hbox{\includegraphics[width=0.5\textwidth]{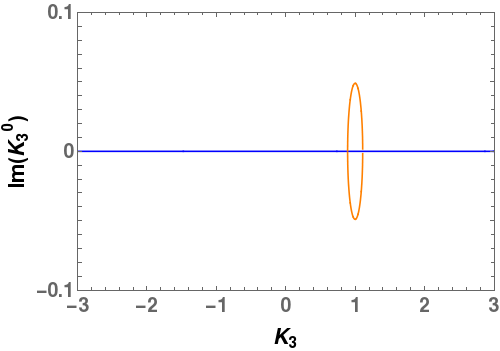}\hspace{0.5cm}\includegraphics[width=0.5\textwidth]{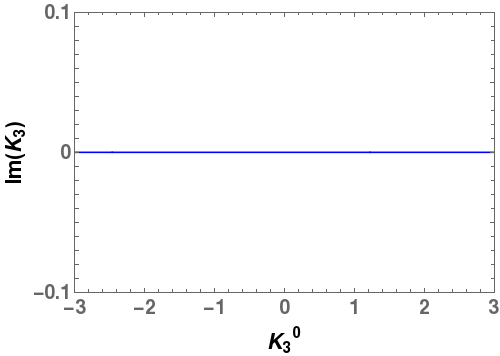}}
\hbox{\includegraphics[width=0.5\textwidth]{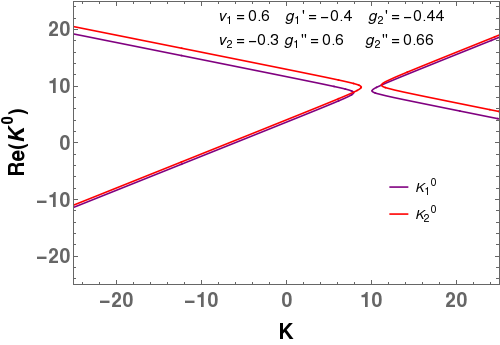}\hspace{0.5cm}\includegraphics[width=0.5\textwidth]{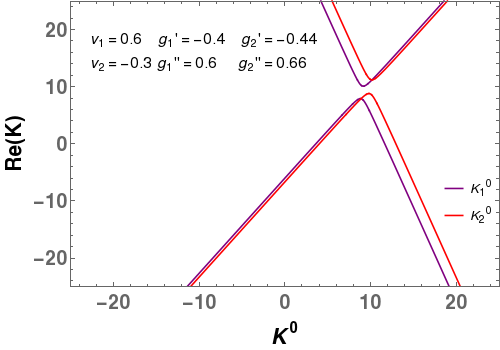}}
\hbox{\includegraphics[width=0.5\textwidth]{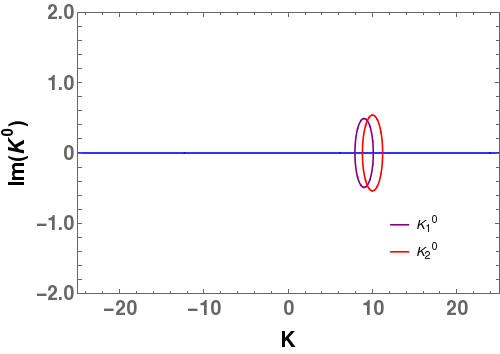}\hspace{0.5cm}\includegraphics[width=0.5\textwidth]{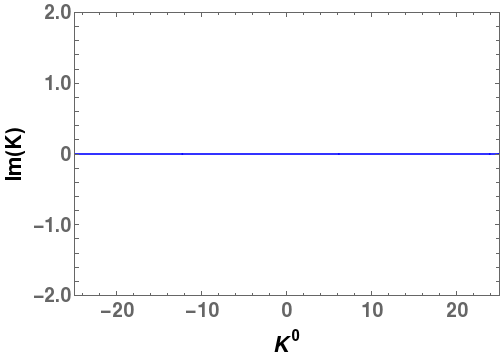}}
\caption{Example of solutions of the dispersion relation (4.12) for an absolutely unstable case.  The upper two panels are for j=3 and lower two panels for j=1,2.  Here $v_1=0.6,v_2=-0.3$. 
$g'_3=-0.04,g''_3=0.06$ and $g'_1=-0.4,g''_1=0.6$, $g'_2=-0.44,g''_2=0.66$.} \label{fig:dispersion3}
\end{figure}

Now that the densities are chosen, the angular densities ($v_1, v_2, g_j'\,g''_j$) has to be fixed for our numerical example. In reference, \cite{Capozzi:2017gqd} it is 
shown that the angular distribution can create 4 different kind of instabilities namely the completely stable, damped stable, convectively unstable and absolutely unstable. In
the following, we take the same angular distributions to generate the different kind of instabilities for our three flavor example. To keep the example simple we consider the nature of the instability to be the same for all the three off diagonal elements, i.e, same for the $\nu_e-\nu_\mu$, $\nu_e-\nu_\tau$ and $\nu_\mu-\nu_\tau$ sectors. The stability properties of the system can be analyzed by studying the conditions which lead to imaginary parts for $K_j^0$ or $K_j$ in the dispersion relation in equations \ref{eq: disprelation}. The presence of instability indicates the possibility of flavor conversion. We plot the real and imaginary parts of $K_j^0$ vs real values of $K_j$ and vice versa in the figures \ref{fig:dispersion2} and \ref{fig:dispersion3}. The gap in $K_j^0(K_j)$ corresponds to the presence of it's complex values in that particular range of real $K_j(K_j^0)$ indicating instability.

The 4 different kind of instabilities mentioned above arises by choosing different numerical values of the parameters $v_1$, $v_2$ and $g_j'$, $g_j''$. Based on these values, the solutions of the dispersion relation are real or complex which in turn decides the type of instability which will arise. This classification of instabilities is mainly dependent on the relative sign of $g_j'$ and $g_j''$ which results in two different situations. The transition from stability to instability requires a change in sign (positive to negative) of the R.H.S of equation \ref{eq: disprelation}. If $g_j'g_j''<0$ i.e. the two terms in equation \ref{eq:2beamangdist} have opposite sign, then the system becomes unstable and $g_j' g_j''>0$ leads to the stable cases. The stable cases include the completely and damped stable ones while the unstable ones can be convective and absolutely unstable. In addition to this, the sign of the product of $v_1$ and $v_2$ is also an important factor for further classifying these cases. $v_1v_2>0$ results in completely stable and convective unstable cases whereas  $v_1v_2<0$ are for damped stable and absolutely unstable scenarios. Taking into account the above conditions, we choose the values of the parameters and plot the solutions of the dispersion relation for the two unstable cases in figures \ref{fig:dispersion2} and \ref{fig:dispersion3}.

The three sectors i.e., $j=1, 2, 3$ show similar behaviour as in the 2 flavor case \cite{Capozzi:2017gqd} for all the four kinds of instabilities. The only difference is that the region of instability is shifted due to the presence of non zero matter terms. However, we are more interested in the unstable cases giving an insight to the possibility of flavor conversions, so the two unstable cases, convectively unstable and absolutely unstable, are explained in detail with plots.
For the stable cases we do verify both the completely stable ( $v_1v_2>0$ and $g_3'g_3''>0$, $v_1=0.7$, $v_2=0.2$, $g_3'=0.04$ and $g_3''=0.06$) and the damped stable ($v_1v_2<0$ but $g_3'g_3''>0$, $v_1=0.6$, $v_2=-0.3$, $g_3'=0.04$ and $g_3''=0.06$) cases for j=3. They mimic the two flavor results with the matter shifting. Assuming the same angular distribution for the other two ($j=1,2$) modes for the stable cases, we find the same behaviour.

The convective instability is shown in figure \ref{fig:dispersion2} for all the j values, i.e., $j=1$, 2 (lower two panels) and $j=3$ (upper two panels), corresponding to all the three sectors (e-$\mu$, e-$\tau$ and $\mu-\tau$) respectively. It is called convective as in this case the perturbation decays locally, but blows up elsewhere as it moves away. We can see that there is a gap in both $K_j^0$ and $K_j$. Both $K_j^0$ and $K_j$ develops imaginary part for a particular range of real values of each other. Here, $v_1v_2>0$ and $g_j'g_j''<0$. In particular, we take $v_1=0.7$, $v_2=0.2$, $g_1'=-0.4$, $g_1''=0.6$, $g_2'=-0.44$, $g_2''=0.66$, $g_3'=-0.04$ and $g_3''=0.06$.

Similarly, the absolute instability case is shown in figure \ref{fig:dispersion3}  for all the three sectors. In this case, the perturbation grows ``on site" and around and thus the name absolutely unstable. Here, the imaginary part arises only in $K_j^0$ for real values of $K_j$. There is a relative change in sign in both $v_1$, $v_2$ as well as $g_j'$, $g_j''$. We take the values as $v_1=0.6$, $v_2=-0.3$, $g_1'=-0.4$, $g_1''=0.6$, $g_2'=-0.44$, $g_2''=0.66$, $g_3'=-0.04$ and $g_3''=0.06$.

One interesting feature which is evident from the lower two panels of figures \ref{fig:dispersion2} and \ref{fig:dispersion3} is that the range of values of $K_j^0$ for which $K_j$ is imaginary and vice versa for the two different sectors are shifted with respect to each other. This means that instability gets shifted. This signifies that the modes unstable for one sector (say $\nu_e-\nu_\mu$) may turn out to be stable for the other (say $\nu_\mu-\nu_\tau$). Moreover, some parts of the two ranges are also overlapping,  indicating modes which are unstable in both the cases. Note that in our example, we choose $\lambda$'s to be of similar order and in units of $\lambda_3\sim1$. Large values of $\lambda$'s will make this effect of different homogeneous modes more pronounce. However, the fast growth of the instabilities may act against these effects. It would be interesting to see in a nonlinear analysis if the presence of matter will have any dominant effect or not. Also our assumption of all the three sectors having same kind of instability may not hold in a realistic situation. Thus, the effect of these different kind of instabilities on each other would be another interesting aspect. 

In particular, for the two flavor picture the fast conversions are only governed by the `electron lepton number' or ELN ($\nu_e-\bar\nu_e$ sector). Thus, $G_{1,v}$ is solely dependent on $\nu_e-\bar\nu_e$. The fast oscillations are connected to the `crossings' in the ELN spectra.
The size of the crossings are also important. The impact of the fast conversions on SN dynamics is one crucial factor \cite{Azari:2019jvr}. Our analysis shows that in the three flavor picture it is not only ELN but in principle all the three flavors are important. It may 
lead to the curious scenario when the ELN crossing is small or absent, however the $\mu$LN or $\tau$LN contribute to the effective lepton crossing. In the non-linear evolution, the instability in one off-diagonal element will influence the growth of instability in the other sectors. Thus the growth of $S_3$ governed by crossings in $\mu$LN/$\tau$LN can impact the instability growth in $S_1$ which is otherwise only dependent on ELN in the two flavor scenario. Given the recent SN simulations \cite{Bollig:2017lki} with $\mu$ being produced in the late accretion phase results in larger difference of fluxes of different flavors than the previous results. It would be interesting to see the sizes of the $\mu$LN or $\tau$LN in these SN simulations and their possible impact on the fast conversions in a SN environment.

\section{Discussion and Conclusion}
\label{sec:conclusion}
The phenomena of self induced flavor conversions in the regions of dense neutrinos (stellar core) have been previously studied using the method of linear stability analysis in the 2 flavor scenario. Here, we have studied the evolution of the system considering all the three flavors of the neutrinos. Considering that the evolution of the matrix of the densities, $\rho$ we obtained three linearized equations of motion, two more compared to the two flavor scenario. We studied these equations for both the slow and the fast instabilities. For the slow oscillations, assuming the system to be isotropic and taking into account only the homogeneous modes, we studied the stationary solutions, i.e., only the spatial evolution. This simple system is similar to the previous three flavor studies done for the dense neutrino problems. Our linear analysis results are in excellent qualitative agreement with these previous non linear three flavor studies. Further, for the fast conversions we did a generalized analysis. We adopted the  dispersion relation approach to study both the temporal as well as spatial evolution. This is the first time the three flavor effects on the fast oscillations are considered. The addition of the third flavor shows possibilities impacting the rate of growth of the instabilities when compared to the two flavor case. Given the suitable conditions, addition of the third flavor may slow or speed up the
instabilities and in turn may affect the flavor conversions.

In the three flavor analysis, the assumption of different distributions for all the three flavors of neutrinos lead to the non trivial effect of the third flavor. We found three linearized equations of motion corresponding to the three off diagonal elements $S_1$, $S_2$ and $S_3$ describing $\nu_e-\nu_\mu$, $\nu_e-\nu_\tau$ and $\nu_\mu$-$\nu_\tau$ sectors respectively. We found that the minimum requirements to reduce the three flavor evolution to an effective two flavor one are $f_{\nu_{\mu}}=f_{\bar{\nu}_{\mu}}$ and $f_{\nu_{\tau}}=f_{\bar{\nu}_{\tau}}$. This is clearly different that the present  assumptions on fluxes, i.e, $f_{\nu_{\mu}}=f_{\nu_{\tau}}$ and $f_{\bar{\nu}_{\mu}}=f_{\bar{\nu}_{\tau}}$ to generate the effective two flavor description.

For the three flavor linearized stability analysis, we found instabilities for both the mass orderings. This is similar to the two flavor studies except that here, the instabilities may remain present corresponding to the $\nu_\mu$-$\nu_\tau$ sector as well. We obtained all the three different kinds  of instabilities namely Bimodal (BM), Multi Zenith Angle (MZA) and Multi Azimuthal Angle (MZA). The BM happens for positive $\mu$, i.e., inverted mass ordering and the MZA and MAA for negative $\mu$, i.e., normal mass ordering. 

The overall matter effect defined in our framework includes the contributions of both the charged leptons and the corresponding neutrinos. It consists of both the current as well as the normal matter term. So in general, the effects from both would be important for the study of the three flavor evolution. We have assumed isotropy for the linearized stability analysis of the slow conversions and the currents have been neglected. However, both the contributions have been taken into account for the dispersion picture of the fast conversions.

According to the previous effective two flavor studies, the onset of the growth of the instability in the solar sector is found to be at a later radius compared to that of the atmospheric sector, however the onset of the solar sector may get faster in the three flavor analysis. Our 3 flavor linearized stability analysis showed that the onset of the growth of the third off-diagonal term ($\nu_\mu$-$\nu_\tau$) occurred at a larger value of $\mu$ i.e., at a lower radius. In the non linear scenario the three evolution equations are coupled to each other, thus the early growth of the third off diagonal can affect (speed up) the other off diagonals and may speed up the solar sector conversions. This is in perfect agreement with the previous three flavor nonlinear studies. We performed the slow conversion stability analysis without considering the ordinary matter effects. However, the consideration of non-zero matter could have substantial effect on the evolution. 
Note, the recent study \cite{Doring:2019axc} involving  two simple models of colliding and intersecting neutrino beams, where the slow onset of the solar sector compared to the atmospheric one was discussed. Our study identified all the different instabilities involved and demonstrated that the this slow onset is happening for all the cases BM, MAA and MZA instabilities. Indeed, the speed up due to the $\nu_\mu - \nu_\tau$ sector would be true to all the three instabilities.

In the dispersion picture for the fast oscillations, we obtained three dispersion relations for the three flavor scenario. Here also $f_{\nu_{\mu}}=f_{\bar{\nu}_{\mu}}$ and $f_{\nu_{\tau}}=f_{\bar{\nu}_{\tau}}$ are the conditions to get back the effective two flavor picture. For the three flavor case we considered the simplest two beam model where each beam consists of all the three flavors of neutrinos. Depending on the values of the angular densities chosen, we obtained four kinds of instabilities, namely the completely stable, damped stable, convectively unstable and absolutely unstable corresponding to each of the off diagonal element. For simplicity, we assumed same angular distributions for all the three sectors resulting in the presence of similar kind of instabilities for all of them. However, the different angular distributions may result in different stability picture in different flavors. In the non-linear evolution this might result in an interesting impact, as in the non linear picture the instabitiy in one flavor sector may have strong influence on the evolution of another flavor sectors.
This becomes interesting when fluxes for the three flavors of neutrinos and their respective antineutrinos are different \cite{Bollig:2017lki}. It would be crucial to understand the influence of these flux differences in the context of these `three flavor speed up (or slow down!)' in the non-linear evolution.

 In the earlier studies involving the dispersion relations, the matter term, being a constant is often rotated away. Similarly, in case of three flavor analysis also, we have constant matter terms but they are different for the three sectors. As a result, they cannot be rotated away simultaneously. The presence of this term shows a very interesting result. It indicates that the modes unstable for some $\lambda$ may not be the same for the others i.e, the effective `homogeneous' mode for one sector may be `inhomogeneous' for the other sectors. However, in the context of fast oscillations these instabilities are growing  at an enormous rate. Hence negligible `cascading' between the modes \cite{Mirizzi:2015fva} may reduce this effect. In particular, for our numerical example the separation of these modes were reasonably small. It would be interesting to see the influence of these effects in future non linear simulations.

\section*{Acknowledgments}

Sovan Chakraborty acknowledges the support of the Max Planck India Mobility Grant from the Max Planck Society, supporting the visit and stay at MPP during the project.
The authors acknowledge Mr. Supriya Pan who was involved in the very early stage of the project and Rasmus S. L. Hansen for initial discussion on the work. This work has received funding/support from the European Union’s Horizon 2020 research and innovation programme through the InvisiblesPlus RISE under the Marie Skłodowska-Curie grant agreement No 690575 and through the Elusives ITN  under the Marie Skłodowska -Curie grant agreement No 674896.

\begingroup\raggedright

\endgroup

\end{document}